\documentclass[
    ,final            
  ,draft            
  ]
  {aipproc}

\layoutstyle{6x9}

\begin{document}

\title 
      [The chemical evolution of the Milky Way: from light to heavy elements]
      {The chemical evolution of the Milky Way : from light to heavy elements}

\classification{}
\keywords{}

\author{Francesca Matteucci}{
  address={Astronomy Department, Trieste University, Via G.B. Tiepolo, 11,
           34124, Trieste, Italy},
  email={matteucci@ts.astro.it}
  thanks={}
}


\begin{abstract}
We present results for the chemical evolution of the Milky Way including predictions for elements from Deuterium to Europium. 
A comparison with the most accurate and recent data allows us to draw important conclusions on stellar nucleosynthesis processes 
as well as on mechanisms of galaxy formation.
\end{abstract}

\maketitle

\section{Introduction}

The Milky Way is the best studied system and a great deal of accurate 
data such as high resolution spectroscopy relative to the solar neighbourhood 
have recently accumulated. Detailed chemical evolution models  able to follow in detail the evolution of the abundances of several 
chemical species in the interstellar gas are also available and by means of them we can compare theory with observations and infer 
many important constraints. In particular, by adopting different set of stellar yields for stars of all masses we can test the 
stellar nucleosynthesis and the origin of elements. Particularly important are the plots relative to abundance ratios, such as [X/Fe], 
as functions of stellar metallicity, measured by the quantity [Fe/H]. Recently, Cayrel et al. (2004) have provided a sample of 
abundances for extremely metal poor stars in the Galactic halo, in particular for stars with metallicity as low as [Fe/H]=-4.0dex. 
The elements studied are mainly $\alpha$-elements (O, Mg, Si, Ca, S, Ti), Fe-peak elements and heavier.Abundances for C and N in 
extremely metal poor halo stars 
were also measured (Spite et al. 2005; Israelian et al. 2004). All of these data represent an invaluable 
template for testing chemical evolution models.
Here we test the two-infall model (Chiappini et al. 1997) for the chemical evolution of the Milky Way on a large sample of data including the abundances discussed above plus the abundance trends recently derived for light elements such as 
deuterium.
The main ingredients of the chemical evolution model are: i) the initial conditions, ii) the star formation history, iii) the initial mass function (IMF), iv) the stellar nucleosynthesis, v) the rate of gas accretion.
In the two-infall model the main assumptions are that the halo and thick disk formed by gas accretion on a relatively short timescale (1-2 Gyr), whereas the thin disk formed by means of an independent accretion episode on longer timescales increasing with galactocentric distances. In particular, the timescale for the formation of the solar neighbourhood is 7-8 Gyr, as imposed by the stellar metallicity distribution.
The assumed rate of gas accretion (primordial gas) is:
\begin{equation}
IR= A(R) e^{-t/ \tau_{H}(R)}+ B(R) e^{-(t-t_{max})/ \tau_{D}(R)}
\end{equation}

with the timescale for the first infall episode $\tau_H(R)$= 2 Gyr for the halo, and the timescale for the disk infall
$\tau_D(R)$ linearly increasing with R which implies an ``Inside-Out Model''.
The time $t_{max}$ is the time of the maximum infall on the thin disk.
\par
The star formation rate is: 
\begin{equation}
SFR= \nu \sigma_{tot}^{k_2} \sigma_{gas}^{k_1}
\end{equation}
with
$k_1=1.5$, $k_2=0.5$, and $\sigma_{tot}$ being the total surface mass density 
and $\sigma_{gas}$ the surface gas density.
A threshold density for the SFR of $7M_{\odot}pc^{-2}$ is also assumed. 
This threshold regulates the star formation regime producing an oscillatory 
behaviour for the SFR when the gas density is close to its value (see Figure 1).

The assumed IMF is multi-slope ($x_1$, $x_2$, ..) and follows the 
prescriptions of Scalo (1986).

\subsection{Element production}

The element production in stars can be summarized as follows:
\begin{itemize}
\item Low \& intermediate mass stars ($0.8 \le M/M_{\odot} \le 8$)
produce mainly $^{4}He$, $^{12}C$, $^{14}N$ and heavy s-process elements
(Ba, Y, Sr) (in stars with masses from 1 to 3$M_{\odot}$).

\item 
Type II SNe ($M >8 M_{\odot}$) produce mainly $\alpha$-elements 
(O, Ne, Mg, Si, S, Ca), part of Fe and other Fe-peak elements as well as 
r-process 
elements (Eu, Ba). 

\item Type Ia SNe produce mainly Fe-peak elements ($\sim 0.6 - 0.7 
M_{\odot}$ of $^{56}Fe$ per supernova).

\item Novae can possibly be important in producing $^{7}Li$, $^{13}C$, 
$^{15}N$, $^{17}O$. $^{7}Li$ can be produced also in SNII, 
red giant, asymptotic giant branch stars and cosmic rays.

\item Deuterium is only destroyed in stars,  whereas $^{3}He$ is destroyed 
but also produced in stars from 1 to 3$M_{\odot}$.

\end{itemize}

\subsubsection{Type Ia SN rates}
The type Ia SNe are the most important Fe producers in the universe and 
represent a fundamental ingredients in galactic 
chemical evolution models.
The most popular scenarios for the progenitors of Type Ia SNe can be 
summarized as follows:
\begin{itemize}
\item {\bf Single degenerate scenario}. This is the  
classical scenario of Whelan and Iben (1973), namely C-deflagration in
a C-O white dwarf reaching the Chandrasekhar mass, $M_{Ch}$,  
after accreting material from a
companion which can be either a red giant or a Main Sequence star.In this case 
the maximum mass allowed for both the primary and secondary star 
in the binary system is 8$M_{\odot}$ and consequently the minimum timescale 
for the occurrence of type Ia SNe 
is $t_{SNIa_{min}}$=0.03 Gyr 
(Greggio \& Renzini 1983; Matteucci \& Recchi, 2001).

\item {\bf Double Degenerate scenario}.This is the
merging of two C-O white dwarfs, due to gravitational wave radiation,
which explode by C-deflagration when the $M_{Ch}$ is reached (Iben
\& Tutukov 1984)The minimum timescale for the occurrence of such 
systems is given by the lifetime of a 8$M_{\odot}$, as above, plus the 
gravitational time delay $t_{SNIa_{min}}=0.03 +\Delta t_{grav}$=0.03+ 0.15Gyr 
(see Tornamb\` e \& Matteucci, 1986).

\item A more recent model by Hachisu et al. (1999) 
is based on the single degenerate scenario but with a
metallicity effect. In particular, no type Ia system can form 
if the metallicity of the progenitor stars is lower than [Fe/H] $= -1.0$.
This model suggests also that only systems with secondary stars with masses 
lower than 2.6$M_{\odot}$ can be accepted. This leads to a minimum 
time for the occurrence of type Ia SNe no shorter than 
$t_{SNIa_{min}} = 0.33$ Gyr, to which one should add the metallicity delay, 
namely the time taken by the chemical enrichment process to reach a 
metallicity of [Fe/H]=-1. 

\end{itemize}

It is important to know the minimum timescale for the occurrence of 
type Ia SNe as well as the 
time for the maximun in the type Ia SN rate, which depends not only upon the 
SN progenitor model but also upon the SFR.

\begin{figure}
\includegraphics[height=.3\textheight]{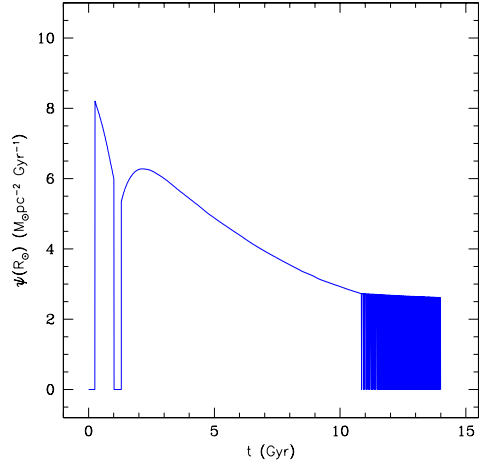}
  \caption{The star formation rate as predicted by the two-infall model by 
Chiappini et al. (1997) for the solar neighbourhood. 
Note the oscillatory behaviour at late times in the thin disk evolution which 
is a consequence of the assumed star formation threshold.}
\end{figure}

\subsection{Results}

\paragraph{The G-dwarf metallicity distribution}

In Figure 2 we present the predicted G-dwarf metallicity distribution compared 
with the observed distributions. The good fit to the data is obtained by means 
of a timescale for the formation of the thin disk in the solar 
vicinity of 7-8 Gyr. A much lower timescale would produce a very bad fit to 
the observed data since it would move the peak of the distribution toward too 
low metallicities (see Kotoneva et al. 2000).
The stellar metallicity distribution is, in fact, very sensitive to the star 
formation history which in turn 
depends strongly on the rate of gas accretion.

\begin{figure}
  \includegraphics[height=.3\textheight]{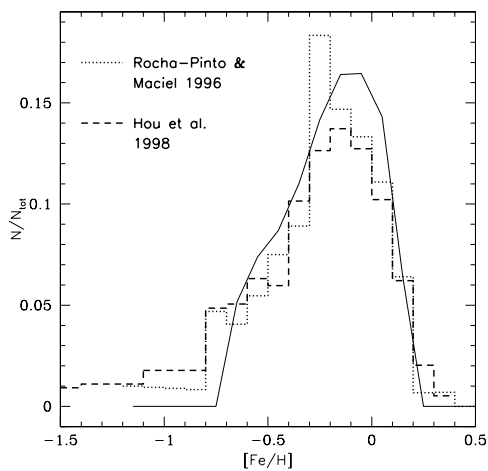}
  \caption{The continuous line represents the predictions of the two-infall 
model with a timescale of disk formation of 8 Gyr.The histograms represent the 
observational data as indicated in the figure.}
\end{figure}

\paragraph{Abundance ratios versus iron abundance}
In Figures 3 and 4 we show the predictions relative to the [$\alpha$/Fe] ratios
versus [Fe/H] compared with a complete set of data including 
Cayrel et al.'s (2004) data, as described in Fran\c cois et al. (2004), who 
tested different sets of yields on the data.
Their conclusion was that the yields of massive stars by Woosley \& Weaver 
(1995, WW95) and those of Iwamoto et al. (1999, I99) for
type Ia SNe are the best to reproduce the observations altough some variations 
are required in the yields of some elements such as 
Mg, Zn and some Fe-peak elements.
The good fits shown in Figures 3 and 4 were obtained by including these 
variations.

\begin{figure}
  \includegraphics[height=.3\textheight]{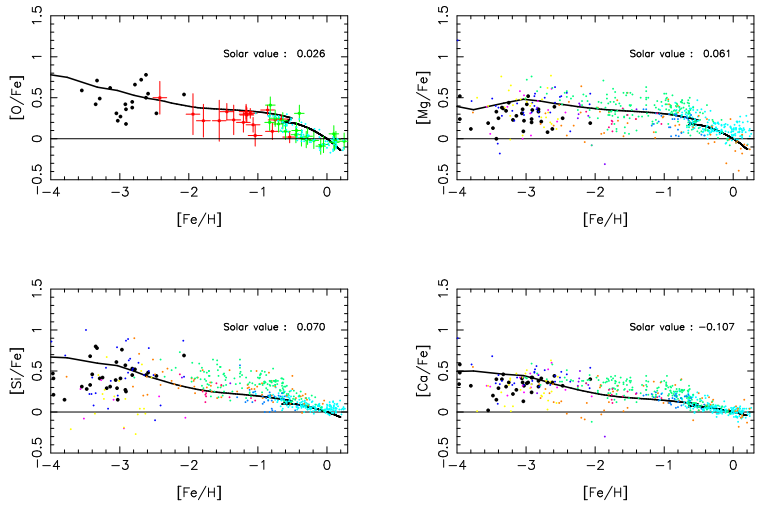}
  \caption{Predicted and observed [$\alpha$/Fe] ratios for a large range of 
metallicities. The model predictions are 
from Fran\c cois et al. (2004) and are normalized to the predicted solar 
abundances. In the top right of each figure 
we indicate our predicted solar ratios, all in good agreement with the 
observed ones([$\alpha$/Fe]=0.0). The black 
dots represent the data from Cayrel et al. (2004).  For the other sources of 
data see Fran\c cois et al. (2004).}
\end{figure}

\begin{figure}
  \includegraphics[height=.3\textheight]{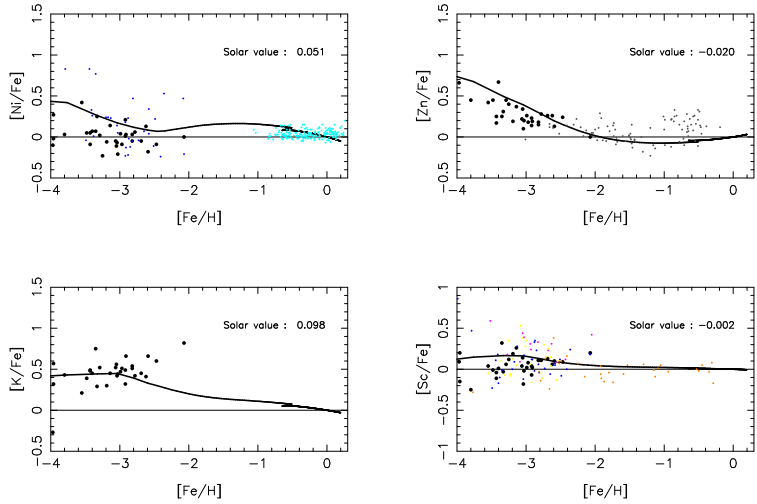}
  \caption{The same as in Figure 3 for other chemical species. The predictions 
are from Fran\c cois et al. (2004). The black dots are the data from 
Cayrel et al. (2004). For the other sources 
of data see Fran\c cois et al. (2004).}
\end{figure}

A remarkable feature of Figures 3 and 4 is the lack of spread at very low 
metallicity, suggesting that the halo of 
the Milky Way was already well mixed at [Fe/H]=-4.0. This should be taken 
into account in 
inhomogeneous models of 
chemical evolution of the halo.

\paragraph{The evolution of light elements}
In Figure 5 we present the evolution of deuterium during the lifetime of the 
Galaxy in the solar vicinity. The 
predictions are from the models of Chiappini et al. (2002) and Tosi (1988). 
The observed values at the time of birth 
of the solar 
system and in the local intertellar medium (ISM) are also shown.
The assumed primordial value for D is the one 
suggested by the WMAP experiment (Spergel et al. 2003). The large spread 
shown for the local abundance of D, as measured by FUSE and HST-GHRS probably 
indicates the effects of D depletion into dust grains.
 It is 
important to notice that, in order to reproduce 
the data, the astration factor for D should have been lower than a factor of 
1.5 (see Romano et al. 2003).

\begin{figure}
  \includegraphics[height=.3\textheight]{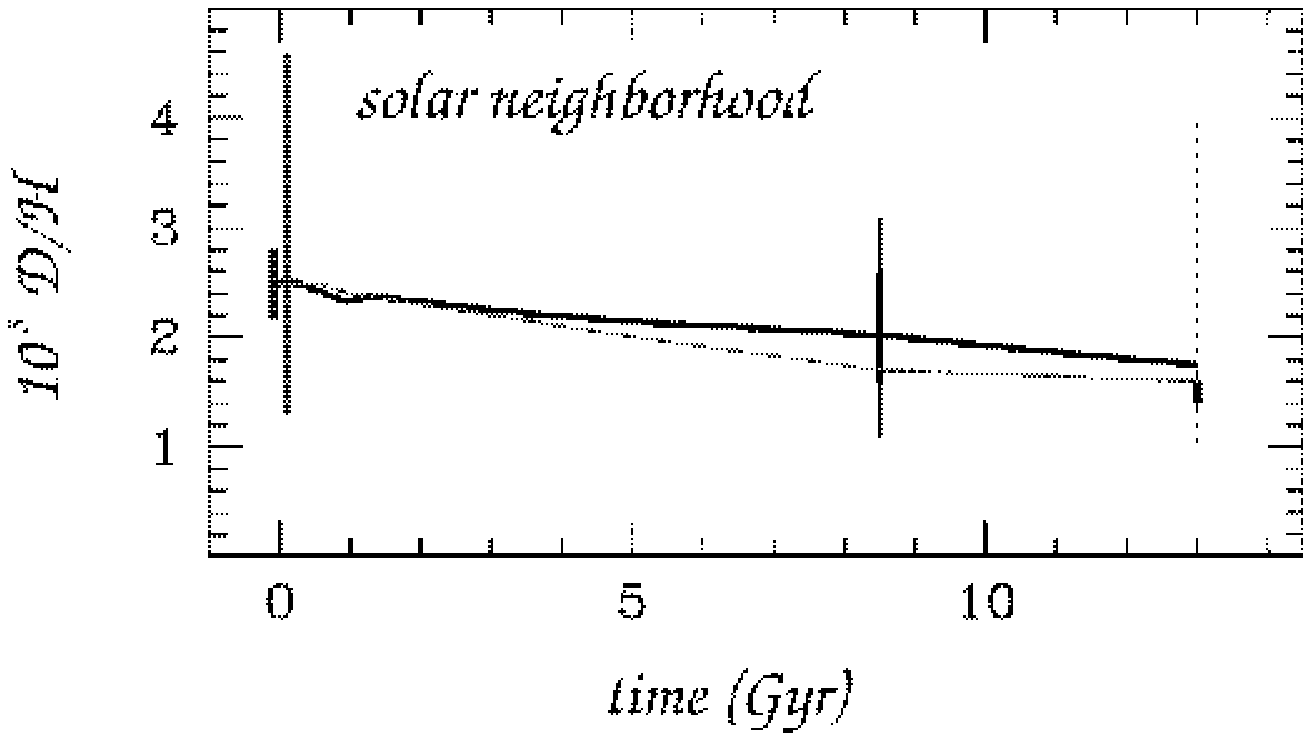}
  \caption{The evolution of deuterium  in the solar vicinity during the 
galactic lifetime. 
The assumed primordial D value is from WMAP (small bar at t=0), the large bar 
at t=0 represents the primordial D as measured from QSO 
absorption spectra. The bar at t=8.5 represents the solar system value from 
Gleiss \& Gloecker (1988).  The D abundance in the local ISM  
(at t=13 Gyr) is from Linsky (1998, small bar) and from FUSE and HST-GHRS 
data (large dotted bar, Vidal-Majar et al. 1998; Moos et al. 2002).
 The thick line represents the 
predictions of the 
two-infall model whereas the thin line 
the predictions of Tosi's (1988) model, as described in Romano et al. (2003).}
\end{figure}

\paragraph{Very heavy elements}

The s- and r- process elements are generally 
produced by neutron capture on Fe seed nuclei.The former 
are formed during the He-burning phase both in low-intermediate  and massive stars, whereas the latter occur in explosive events such as Type II SNe. Recently, Fran\c cois et al. (2006 and this conference) have measured the abundances of several very heavy elements (e.g. Ba and Eu) in extremely metal poor stars of the Milky Way. Previous work on the subject had shown a large spread in the abundance ratios of these elements to iron, especially at low metallicities. This spread is confirmed although is less than before, and is at variance with the lack of spread observed in the other elements shown before. 
Apart from this problem, not yet solved, these diagrams can be very useful to put constraints on the nucleosynthetic origin of these elements. In particular, Cescutti et al. (2006) by adopting the two-infall model predicted the evolution of [Ba/Fe] and [Eu/Fe] versus [Fe/H], as shown in figures 8 and 9. They can well fit the average trend but not the spread at very low metallicities since the model assumes instantaneous mixing. In order to fit the Ba evolution, they assumed that Ba is mainly produced as s-process element in low mass stars but that a fraction of Ba is also produced as an r-process element in stars with masses 10-30$M_{\odot}$. Europium is only an r-process element  produced also in the range 10-30$M_{\odot}$.

 \begin{figure}
  \includegraphics[height=.3\textheight]{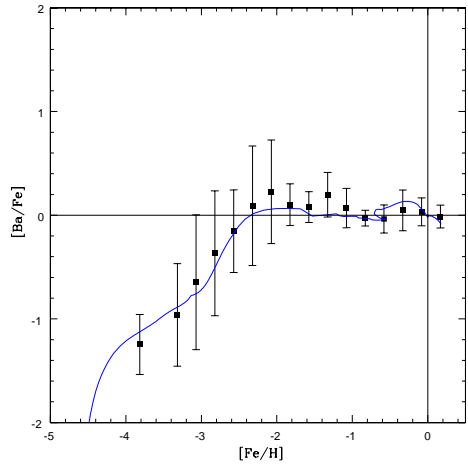}
  \caption{The evolution of Barium in the solar vicinity as predicted by the two-infall model (Cescutti et al. 2006). 
Data are from Fran\c cois et al. (2006)}
\end{figure}

\begin{figure}
  \includegraphics[height=.3\textheight]{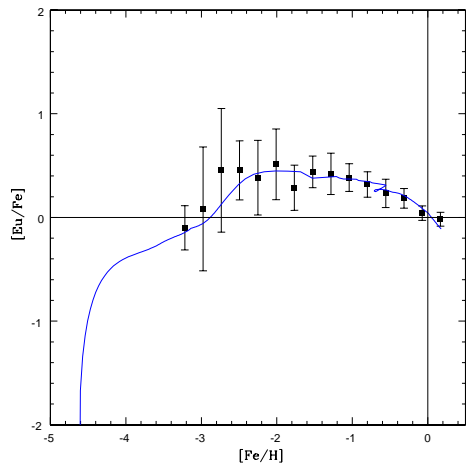}
  \caption{The evolution of Europium in tthe solar vicinity (Cescutti et al. 2006). Data are from Fran\c cois et al. (2006)}
\end{figure}

\section{Conclusions}

Our main conclusion on the origin of the elements and the mechanism of formation of the Milky Way are:
\begin{itemize}

\item The elements $^{12}C$ and $^{14}N$ are mainly produced in low and intermediate 
mass stars ($0.8 \le M/M_{\odot} \le 8$). 
The amounts of primary and secondary N are still uncertain. Nitrogen should be primary in massive stars.

\item
The $\alpha$-elements originate in massive stars: the
nucleosynthesis of O is rather well understood
and there is agreement between different authors.

\item Magnesium is generally underproduced by standard nucleosynthesis models. Starting from the yields of WW95 we suggest that
to fit the data the Mg production should be increased in  stars with $M \le 20 M_{\odot}$ 
and decreased in stars  with $M  > 20 M_{\odot}$.  
Silicon should be only slightly increased in stars with  $M>40M_{\odot}$ (Fran\c cois et al. 2004).

\item Iron originates mostly in type Ia SNe,
and the Fe production from massive stars is still uncertain. 
The metallicity dependent yields of Fe from WW95
overestimate Fe
in stars with   $M < 30M_{\odot}$. It is better to adopt the Fe yields for solar metallicity all over the whole metallicity range. 
The Fe yields of I99 do not show this problem and  they are computed only for solar metallicity (Fran\c cois et al. 2004).

\item The Fe-peak elements: Cr, Mn, Ni, Co. Always starting from WW95 we suggest that the production of Cr and  Mn should be 
increased in stars  with 10-20 $M_{\odot}$.
Cobalt production should be higher in type Ia SNe and lower in 10-20$M_{\odot}$,
Ni production should be lower in type Ia SNe. The type Ia SN yields adopted for comparison are those of I99 (Fran\c cois et al. 2004).

\item In order to reproduce the evolution of Zinc relative to Iron and in particular its overabundance at very low metallicity, new yields from massive stars are required. They should produce more Zn than the standard yields. A good approximation is to use the Zn yields from WW95 for solar metallicity. Moreover, Zn production from type Ia SNe should also be increased (Fran\c cois et al. 2004). 

\item Barium is mainly an s-process elements produced in the 1-3$M_{\odot}$
range, but it must have also an r-component from 10-30$M_{\odot}$ stars. Europium is only
an r-process element formed in the stellar mass range  10-30$M_{\odot}$ (Cescutti et al. 2006).

\item  Deuterium is only destroyed in stars. A very good
agreement with observations is obtained when the primordial abundance $(D/H)_{p}=2.5 \cdot 10^{-5}$
derived from WMAP results is adopted, 
and the astration factor during the galactic lifetime is $< 1.5$ (Romano et al., 2003).

\item In order to fit the G-dwarf metallicity distribution in the solar vicinity, a timescale for the thin disk formation
at the solar galactocentric distance of 7-8 Gyr is required. To reproduce the observed abundance gradients along the galactic disk  
an inside-out formation mechanism with the timescale for disk formation increasing with galactocentric distance should be assumed 
(Matteucci \& Fran\c cois, 1989; Chiappini et al. 2001) .

\end{itemize}




\bibliographystyle{aipprocl}
\bibliography{sample}

\end{document}